\documentclass[a4paper,11pt]{article}
\usepackage{jinstpub} 
\usepackage{lineno}
\usepackage{orcidlink}
\usepackage{tabularx}
\usepackage{graphicx}
\usepackage{subcaption}
\usepackage{siunitx}
\usepackage{xcolor}

\title{\boldmath A 28nm Multiply-Accumulate ASIC Architecture for On-Chip Data Compression in MHz Frame Rate X-ray and Electron Pixel Detectors}

\author{$^{1,2}$Rami Rasheedi\orcidlink{0009-0009-8410-8011}}
\author{$^{1,3}$Nicholas Contini\orcidlink{0009-0006-1918-397X}}
\author{$^{1,2}$Mohamed Adel Gharib\orcidlink{0009-0008-3606-6200}}
\author{$^{1}$Sebastian Strempfer\orcidlink{0000-0001-6235-040X}}
\author{$^{1}$Senthil Gnanasekaran\orcidlink{0009-0003-1637-3354}}
\author{$^{1,2}$Salma Abdelzaher\orcidlink{0009-0003-7242-5665}}
\author{$^{1}$Tejas Guruswamy\orcidlink{0000-0002-4165-6765}}
\author{$^{1}$Kazutomo Yoshii\orcidlink{0000-0003-1904-5383}}
\author{$^{1,4}$Mike Hammer\orcidlink{0000-0003-0395-8225}}
\author{$^{1,4}$Henry Shi\orcidlink{0000-0001-6744-1941}}
\author{$^{4}$Yu-Sheng Chen\orcidlink{0000-0002-7646-7761}}
\author{$^{5}$Lorenzo Rota\orcidlink{0000-0002-5075-210X}}
\author{$^{5}$Dionisio Doering\orcidlink{0000-0001-5182-9044}}
\author{$^{5}$Angelo Dragone\orcidlink{0000-0002-0795-2579}}
\author{$^{1}$Tao Zhou\orcidlink{0000-0002-8093-7666}}
\author{$^{1,4}$Antonino Miceli\orcidlink{0000-0001-5994-5402}}

\affiliation{$^1$Argonne National Laboratory, Lemont, IL, U.S.A.}
\affiliation{$^2$University of Illinois Chicago, Chicago, IL, U.S.A.}
\affiliation{$^3$Ohio State University, Columbus, OH, U.S.A.}
\affiliation{$^4$University of Chicago, Chicago, IL, U.S.A.}
\affiliation{$^5$SLAC National Accelerator Laboratory, Menlo Park, CA, U.S.A.}

\emailAdd{ralsaeed@anl.gov}

\abstract{Modern X-ray detector systems urgently require compact, efficient, and fast data compression schemes to handle the transmission of big data from pixel arrays, enabling frame rates in the MHz regime. In this work, a data compression ASIC that implements a streaming fixed-length lossy compression scheme is introduced and analyzed, proving the feasibility and benefits of on-chip compression. The compression scheme utilizes a vector matrix product logic, which performs a number of floating-point multiplications, additions, and accumulations. The logic is verified, synthesized, and shown to fit in the area resource available for the X-ray detector under study, which comprises 192 \texttimes\ 168 pixels each of 12-bit width, and having a total area of \SI{20}{\mm} \texttimes\ \SI{20}{\mm}, about \SI{2}{\mm} \texttimes\ \SI{20}{\mm} of which are available for the digital logic. Several system architectures, precisions, and compression ratios ranging from 100 to 250 were analyzed to pave the way for on-chip fixed-length compression (e.g., principal component analysis, singular value decomposition) and data reduction (e.g., azimuthal integration) for X-ray and electron detectors.}


\begin{document}
\maketitle
\flushbottom

\section{Introduction}
\label{sec:intro}
X-ray and electron detectors are critical enablers of scientific breakthroughs, powering advances across disciplines from structural biology to materials science, thanks to sophisticated pixel array detectors implemented on application-specific integrated circuits (ASICs)~\cite{EIKENBERRY2003260, DenesSchmitt-2012, HatsuiGraafsma-2015, Graafsma2018, jiang2018electron}. Despite these advances, many scientific detectors still rely on legacy CMOS technology nodes (i.e., above-100 nm CMOS nodes), limiting their speed, integration, and data-handling capabilities. Meanwhile, the demand for larger, faster detectors is driving exponential growth in data volume. Using the most advanced CMOS technology node accessible for scientific pixel detector development, such as \SI{65}{\nm} or \SI{28}{\nm}, the analog-mixed signal components (e.g., analog front-ends, charge-sensitive amplifiers, ADCs) can reach at least \SI{1}{\MHz}~\cite{tate2016high, AGIPD-front-end-2012,DSSC-ASIC-2012,BonnADC-2013,SAR-ADC-SLAC-2024, Falcon-2025}. However, conventional digital readout schemes that transmit all the raw data become a bottleneck as we enter the sub-100 nm CMOS regime. For example, at this frame rate, a 200 \texttimes\ 200 pixel matrix at 12-bit resolution would require 480 Gbps off-chip bandwidth. If we constrain ourselves to the same CMOS technology node for both the analog and digital without invoking 3D vertical integration, an optimistic estimate for the off-chip bandwidth is about $16\times10 = 160\,$Gbps~\cite{Timepix4}. This is an optimistic estimate since achieving 16 lanes at 10 Gbps is a non-trivial engineering challenge (e.g., PCB signal integrity \cite{gharib2025efficient}). In addition, one needs to consider real-world system-level issues beyond the ASIC (e.g., downstream FPGAs I/O configuration, cooling, data management, etc). In addition, the more transceivers there are, the more power there is, and thus reducing transceivers is important from a system and power perspective. To tackle this issue, efficient on-chip data compression is crucial for minimizing data volume prior to serialization and transmission, thus minimizing transmission costs. Our approach shifts into the digital domain early in the readout chain, which allows for faster and more reliable data handling. This enables real-time compression and reduction of data on-chip~\cite{UHR-2023, Rota_2024, RT1-testing-2024, RT2-design-2024}.


In this work, we present a scalable multiply–accumulate architecture implemented in a \SI{28}{\nm} CMOS technology node, which offers an excellent balance of characteristics for mixed-signal integration. This node provides high logic density and high-speed operation for digital circuits, while being less challenging for analog design compared to very deep nodes. The proposed system operates at frame rates of \SI{1}{\MHz}, employing hardware-based data compression to mitigate bandwidth limitations and maximize the performance of pixel detectors in next-generation X-ray and electron imaging applications.


The compression methods utilized in this work are introduced in Section~\ref{sec:Compression_Method}, which includes a comparison of fixed-length versus variable-length encoding. Section~\ref{sec:System_Architecture} provides an overview of the compressor system architecture, highlighting key design blocks, data partitioning, and optimization strategies such as pipelining, logic sharing, and data width management. The RTL implementation is provided in Section~\ref{sec:RTL_design}. Section~\ref{sec:Determining_Necessary_Data_Widths} examines precision requirements and bit-width choices for various datapath components. Section~\ref{sec:Logic_Synthesis} discusses the logic synthesis results and trade-offs associated with both constant and optimized data widths. Finally, Section~\ref{sec:Physical_Implementation} outlines the physical implementation, followed by a discussion in Section~\ref{sec:Discussion}, and concludes the work in Section~\ref{sec:Conclusion}.

\section{Moving from Variable to Fixed Length Compression}
\label{sec:Compression_Method}
Several data compression schemes, such as Zeromask~\cite{Hammer_2021}, sparsification~\cite{Rota_2024}, and Poisson encoding~\cite{huang2021fast, Strempfer_2022} have been studied for on-chip data compression. The Zeromask scheme utilizes the elimination of zero value pixels, and the addition of one more data piece called the metadata, indicating the locations of the pixels, the zero and non-zero pixels. On the other hand, the pixel width zero suppression scheme manipulates the pixel itself, as it adjusts the final data width to the largest pixel in a sequence of pixels, and eliminates all the zeros in the most significant bits of the pixels. Although lossless Zeromask and zero suppression schemes are relatively easy to implement in hardware, their final data width and compression ratio depend on the sample itself; therefore, they are not sample-agnostic, which requires more data handling before sending off-chip. It is worth mentioning that both schemes may also need a bit-shuffling~\cite{masui2015compression} stage to even out the compression ratios across the pixel array, as most of the X-ray image frames have high-value pixels concentrated in the middle, while other regions in the frame contain zero or low-value pixels.

\begin{figure}[]
\centerline{\includegraphics[scale = 0.35]{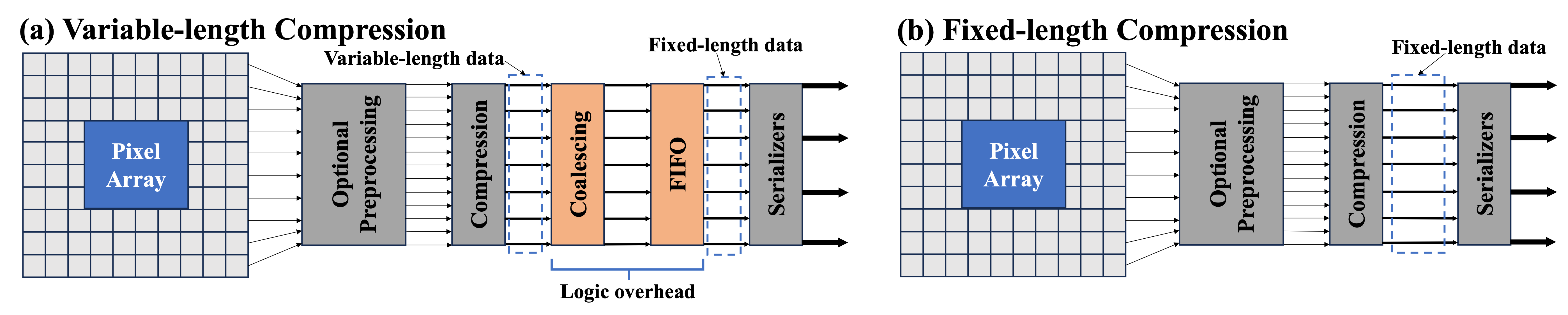}}
\caption{(a) Variable-length compression scheme. (b) Fixed-length compression scheme.}
\label{variablevs.fixed}
\end{figure}

Compression schemes such as Zeromask and sparsification produce variable-length compressed data, meaning the compression ratio varies depending on the characteristics of the data set being compressed~\cite{zhan2008test}. This variability in length introduces additional logic overhead that must be added before the data can be sent to the serializing stage. 
Figure~\ref{variablevs.fixed} illustrates the logic overhead associated with variable-length compression schemes. A coalescing stage is required to gather the variable-length data and pack it into fixed-length segments. The frequency at which these segments are sent is not predetermined, as the input data can vary in length. Therefore, a FIFO (First In, First Out) stage is necessary to ensure that these segments are transmitted to the serializing stage at a consistent frequency. However, we can still overflow if the compression ratio becomes too low for the FIFO to accommodate. Finally, if we consider the case of a multi-ASIC detector system, variable-length compression presents additional challenges. The variability in data rate across a multi-ASIC system can no longer be accounted for by load balancing, limiting the overall frame rate to that of the single ASIC with the worst compression. All of these challenges motivate a transition to fixed-length compression schemes, especially when considering multi-ASIC systems with non-uniform X-ray intensity distributions.

In contrast, fixed-length compression schemes do not require a coalescing stage since the input data is already of a defined length. Additionally, a FIFO stage may not be necessary if both the system and the serializing stage clocks are phase-locked. The predictable nature of fixed-length compression algorithms is attractive for hardware implementations, such as on-chip compression. An emerging fixed-length compression algorithm is one that is based on principal component analysis. Principal component analysis (PCA) is a statistical technique used for dimensionality reduction by transforming correlated variables into a set of uncorrelated principal components. It captures the directions of maximum variance in the data, allowing for efficient representation with fewer components. This makes PCA an effective method for data compression, preserving essential features while reducing storage and computational cost. PCA has been used for 1D data pulse processing application~\cite{PCA-pulses}. 2D PCA-based compression has been explored to X-ray ptychography~\cite{panpan-pca-ptycho,FNAL-PCA}, X-ray photon correlation spectroscopy (XPCS)~\cite{XPCS-SVD}, and X-ray and electron diffraction~\cite{Kutsukake31122024, gladyshev2023lossycompressionelectrondiffraction}. Since PCA compression requires a multiply-accumulate architecture, it has motivated us to develop such an architecture.
\section{Compressor System Architecture}
\label{sec:System_Architecture} 
\begin{figure}[]
\centerline{\includegraphics[scale = 0.3]{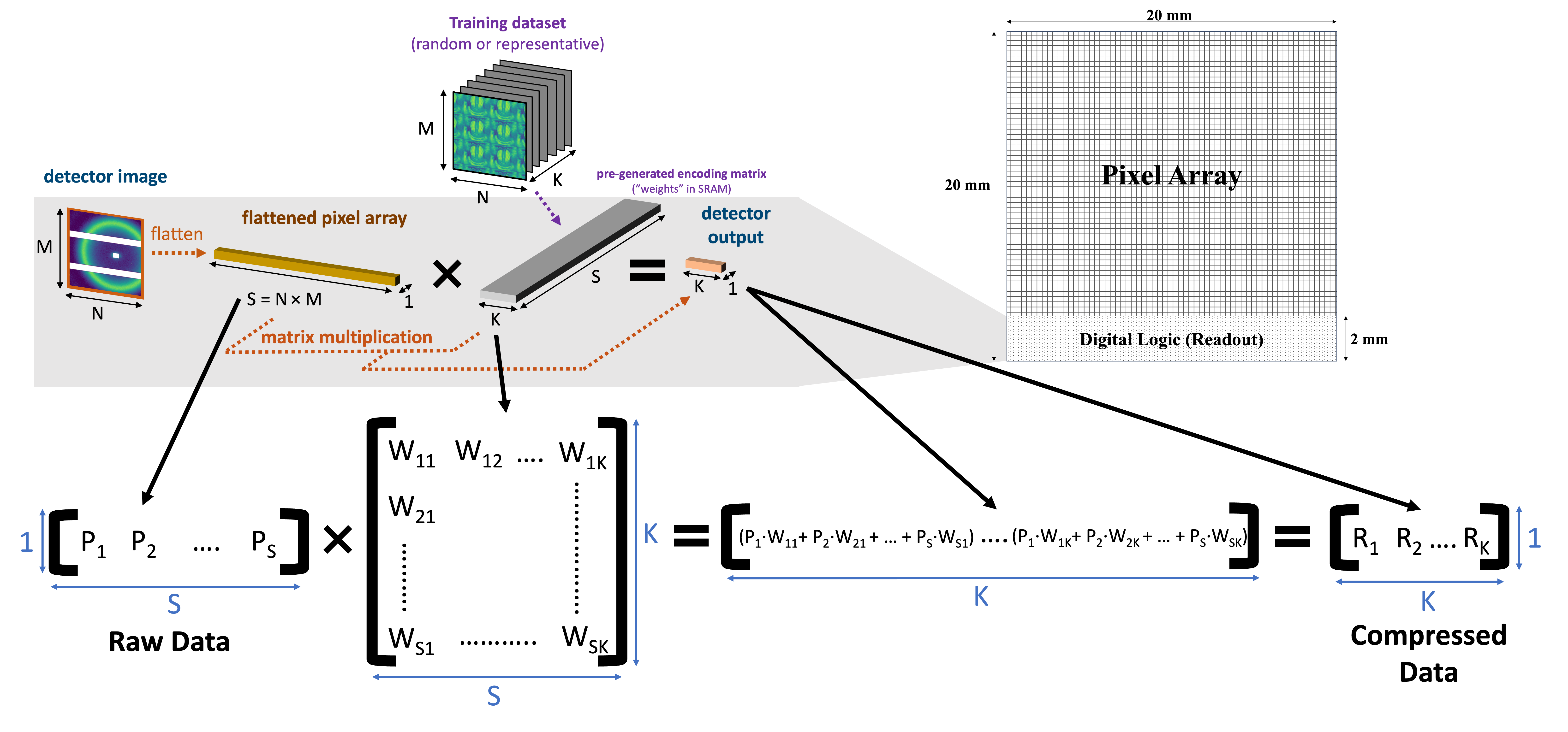}}
\caption{Online matrix multiplication-based compression, where the pixel matrix is flattened out and multiplied by the pre-generated encoding matrix, which is stored on-chip in SRAM. N and M are the number of columns and rows in the pixel array, respectively. The detector ASIC output (not to scale) is compressed to a vector of length  \( K \) -- the number of eigenvectors (principal components) which are kept.}
\label{sparkpix-matmul-figure}
\end{figure}

\begin{figure}[]
\centerline{\includegraphics[scale = 0.3]{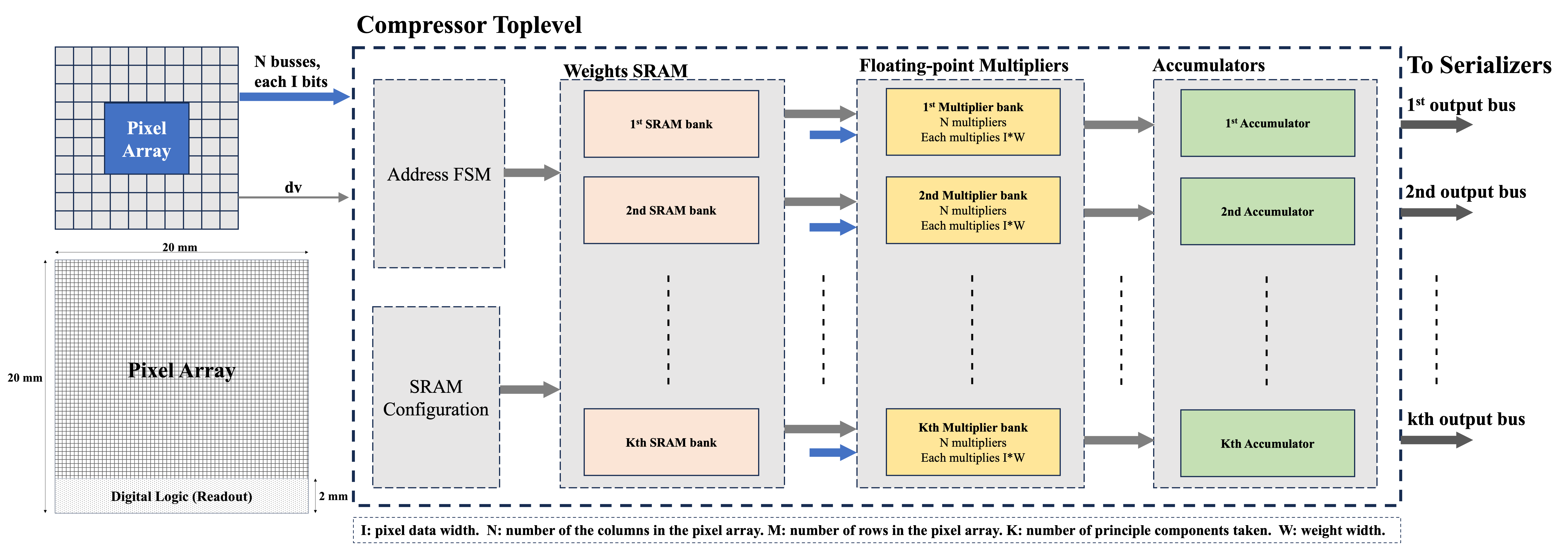}}
\caption{Block diagram of the multiply-accumulate ASIC architecture.}
\label{sys}
\end{figure}

In PCA-based compression, the process begins by using the matrix \( V \) from the singular value decomposition (SVD) of \( X \), which contains the eigenvectors of the correlation matrix \( X^T X \), where \( X \) represents the data to be compressed. In the encoding step, this matrix \( V \) is then multiplied by \( X \). The number of the first \( K \) eigenvectors (principal components) selected from \( V \) determines both the accuracy and the compression ratio~\cite{XPCS-SVD}. A higher value of \( K \) results in a lower compression ratio and greater accuracy, while a lower \( K \) leads to a higher compression ratio but reduced accuracy. In our design, online compression is employed with an encoding matrix that is pre-generated from previously captured data. Because scientific data from a given instrument and sample type are often self-similar, they can be approximated using a common set of eigenvectors. A future publication will describe the general workflow for online compression that can detect drifts and reconfigure weights to handle evolving samples. This approach eliminates the need to wait for the entire set of frames to be delivered, allowing for real-time compression as the frames are processed.
The detector being studied features a pixel array of 192 columns \texttimes\ 168 rows with a \SI{1}{MHz} frame rate, with each pixel value represented using 12 bits. Operating at a frame rate of 1 MHz, the data for each row (comprising 192 pixels) is sent to the compressor, accompanied by a data valid signal every clock cycle. To process the entire frame within 1 microsecond, the compressor must operate at a minimum frequency of 168 MHz. The detector total size is about \SI{20}{\mm} \texttimes\ \SI{20}{\mm}, where \SI{2}{\mm}  \texttimes\ \SI{20}{\mm} one edge of the ASIC (i.e., balcony) is dedicated to the compressor logic. The pixel array area is limited to the analog front-end and analog-to-digital converters. The digitized pixel data is streamed to the balcony for compression. 
As illustrated in Figures~\ref {sparkpix-matmul-figure}, ~\ref {sys}, the compressor primarily performs a matrix multiplication operation, which involves a series of floating-point multiplications, additions, and accumulations. The weights (encoding matrix) are populated based on the applied compression algorithm, which are usually represented using one of the floating-point formats. 

\subsection{Design Blocks}
\subsubsection{Address Finite State Machine}
The address FSM generates the required addresses for the weights stored in SRAM, allowing the system to read the correct weight at the appropriate time and provide it to the floating-point multiplier. Additionally, it produces error signals related to data and frame synchronization in case anything is out of sequence.

\subsubsection{SRAM Configuration}
We anticipate that the pre-generated (i.e., off-chip) encoding weight matrix stored in SRAM will need to be updated dynamically, as continuous learning strategies may be employed \cite{Babu2023}. We have implemented a simple SRAM configuration scheme. The write data comes in the form of a single-bit high-speed serialized stream, which is then organized as write data and write address segments. The SRAM configuration module distributes this write data, along with the corresponding write addresses and write enable signals, to all the SRAMs in the system prior to operation. If we consider the case of K = 192 and 12-bit weights, this equates to \SI{74.3}{Mb}  of SRAM memory. Reconfiguration of all the weights of the SRAMs will take \SI{14.9}{ms}, assuming one \SI{5}{Gbps} serial link dedicated to reconfiguring the SRAMs. This timing does not account for the protocol overhead on the serial link. There may be scenarios when not all the SRAMs need to be updated, so in principle, this reconfiguration time could be decreased. 


\subsubsection{Weights SRAM}
The SRAM is responsible for storing the weights, which can be updated dynamically. The number of SRAM units required depends on several factors, including \( K \), the number of columns in the frame, the frequency of operation, and the width of the weights.
For instance, the number of SRAM units increases linearly with \( K \), assuming all other parameters remain constant. The same linear relationship applies to the number of columns in the frame. However, the number of SRAM units is inversely proportional to the operating frequency. If the frequency is increased by a factor of 4, we can use SRAMs with four times the depth, as we can read four times faster, leading to a reduction in the number of required SRAM units to one quarter.
The relationship between the width of the weights and the number of SRAM units needed is more complex. Typically, if the SRAM width remains constant, a greater number of SRAM units will be required for wider weights. 
The depth of the SRAM is determined by the number of rows in the frame and the frequency of operation, while the width is chosen to maximize area and power efficiency based on the selected technology. Ultimately, the number of SRAM units required to store the weights, along with their respective widths, is influenced by several factors, including the desired compression ratio, precision, available area resources, and the specific compression algorithm used.

\subsubsection{Floating-Point Multipliers}
\begin{figure}[]
\centerline{\includegraphics[scale = 0.33]{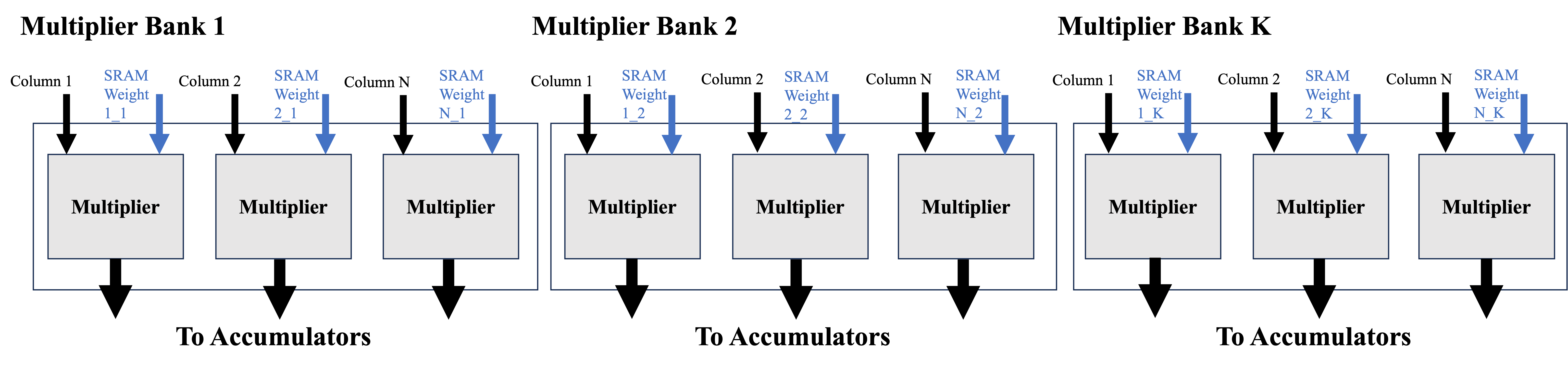}}
\caption{Floating-point multiplication level.}
\label{mult}
\end{figure}

Weights are typically represented using IEEE floating-point formats, such as IEEE FP32 or IEEE FP16. Floating-point multipliers are designed to perform the multiplication operations necessary for matrix multiplication between integer-valued pixel data and the floating-point weights. As illustrated in figure~\ref{mult}, the number of multipliers is equal to \textit{K} \texttimes ~\textit{Number of Columns}, while their size depends on both the pixel data width and the weight width. The data from a single column is multiplied \( K \) times with the corresponding weights stored in the SRAM. After this operation, all the results generated from the multiplier bank are sent to the corresponding accumulator.

\subsubsection{Accumulators}
\begin{figure}[]
\centerline{\includegraphics[scale = 0.4]{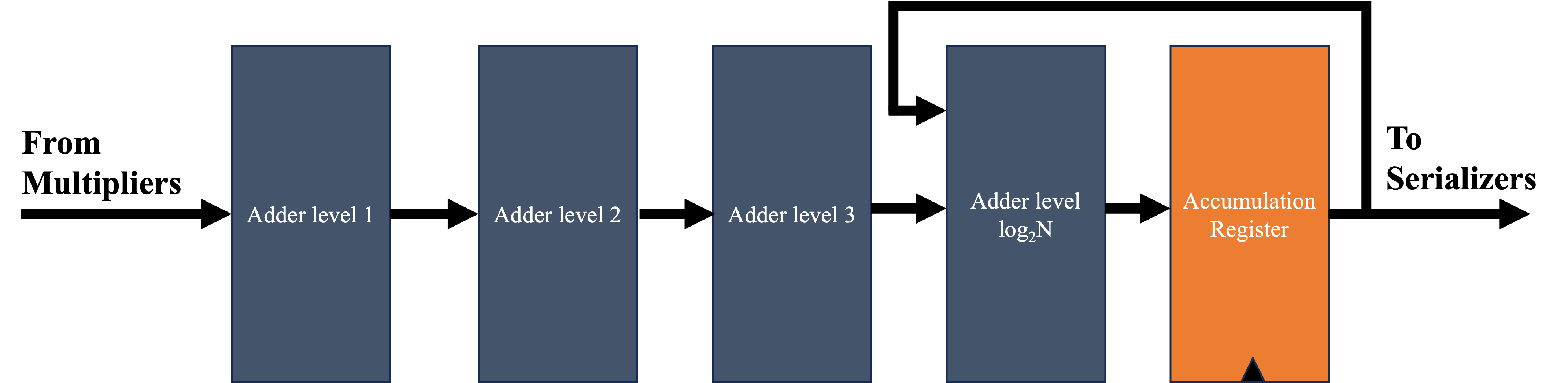}}
\caption{Accumulation level.}
\label{acc}
\end{figure}

Matrix multiplication involves repeated multiply-and-accumulate operations, where each element of the result is computed by summing the products of corresponding elements. These operations are executed using an accumulator stage that follows the multiplier stage. As shown in Figure~\ref{acc}, the accumulator consists of multiple levels of addition. The number of summation levels is determined by the number of columns in the matrix, denoted as \textit{N}. Specifically, the number of addition levels is given by \( \log_2 N \). This is followed by a register stage that stores the summation output, allowing for re-addition to complete the accumulation process. If the compressor block performs matrix multiplication on the entire 192 pixels of the pixel array, the outputs from the accumulators are sent directly to the serialization stage for off-chip transmission. However, if the operation is partitioned, the outputs of the accumulators must be summed again until the data is reduced to the required width before being sent to the serializer.

\subsection{Design Partitioning}
\begin{figure}[]
\centerline{\includegraphics[scale = 0.3]{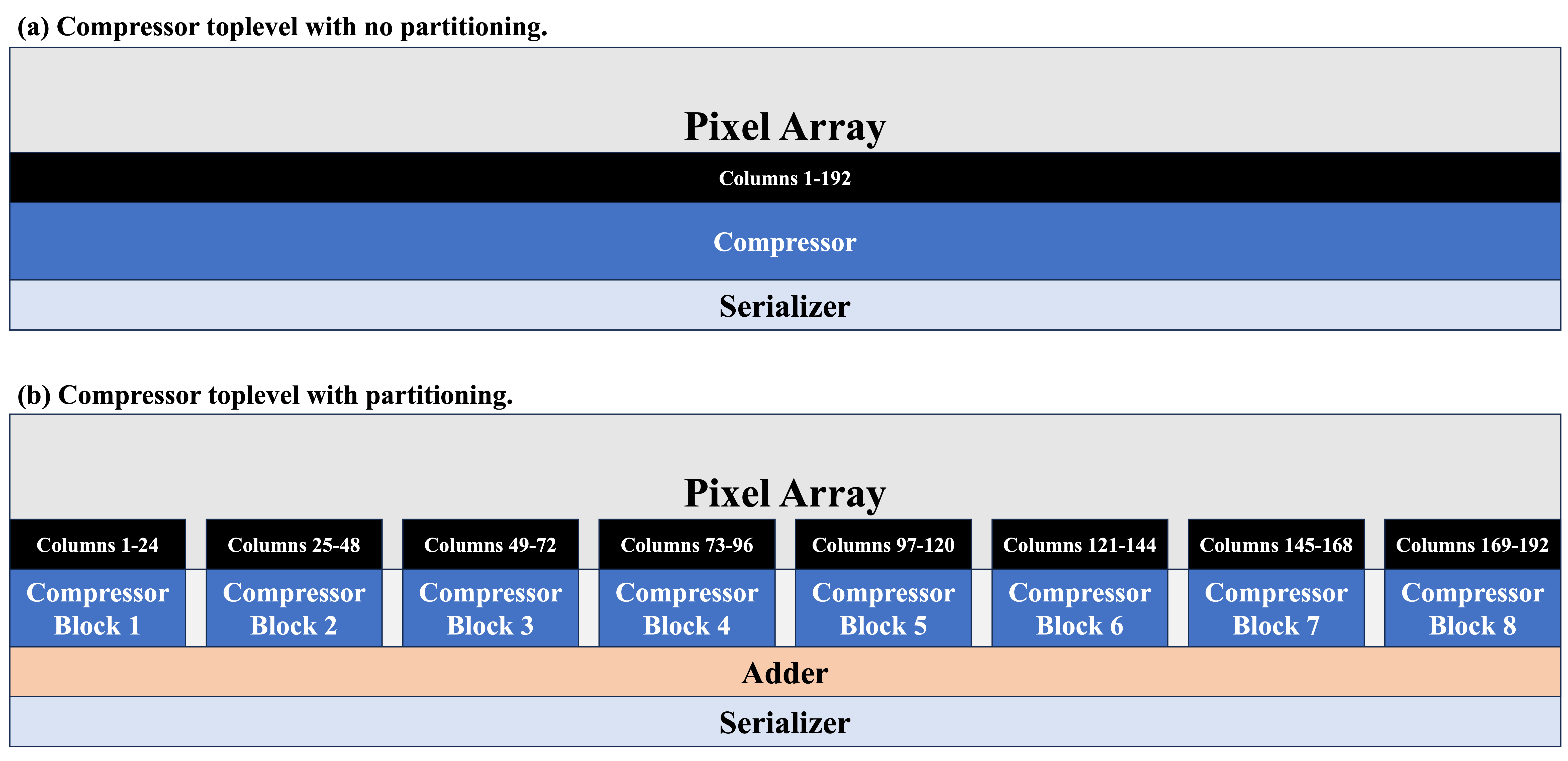}}
\caption{Compressor top-level partitioning.}
\label{part}
\end{figure}

For large values of \( K \), the logic and corresponding SRAM area become significant, leading to inefficiencies in physical implementation and excessively long runtimes. To address this issue, a partitioning scheme should be employed to divide the top-level design into smaller, identical blocks, which facilitates physical implementation.
As illustrated in Figure~\ref{part}, when partitioning is not used, the total number of output buses generated by the accumulators is \( K \). However, when the design is partitioned—such as into 8 blocks—the number of output buses from the accumulators increases to \( 8K \), necessitating the use of additional adder levels after the accumulators to reduce them back to \( K \). It is essential to recognize that design partitioning serves as a method for optimizing both design and runtime performance; it does not influence the design in an algorithmic manner.

\subsection{Design Optimization}
Due to the large size of the design and its high clock frequency, several optimization strategies should be implemented to improve area and timing efficiency. For example, a high number of columns in each partitioning block leads to an increase in the number of adder levels in the accumulators. This subsequently deepens the critical timing path, making it challenging to meet the setup timing at a clock frequency of 168 MHz or more.

\subsubsection{Pipelining}
Pipelining registers are placed along the matrix multiplication pathway, specifically between the multiplier and the accumulator, as well as between the different levels of addition within the accumulator itself. This arrangement shortens the critical path and makes it independent of the number of summation levels, providing greater design flexibility.

\subsubsection{Logic Sharing}
To conserve area, the matrix multiplication logic can be shared among multiple column groups rather than being dedicated to just one. This can be achieved by doubling or quadrupling the frequency, allowing the logic to serve two or four groups of columns at the same frame rate. It is required to insert an additional level of multiplexers in the compressor block to switch between the sharing column groups. However, it is worth noting that while this method reduces the logic area, the SRAM area remains constant with a fixed value of \( K \). 

\begin{figure}[]
\centerline{\includegraphics[scale = 0.39]{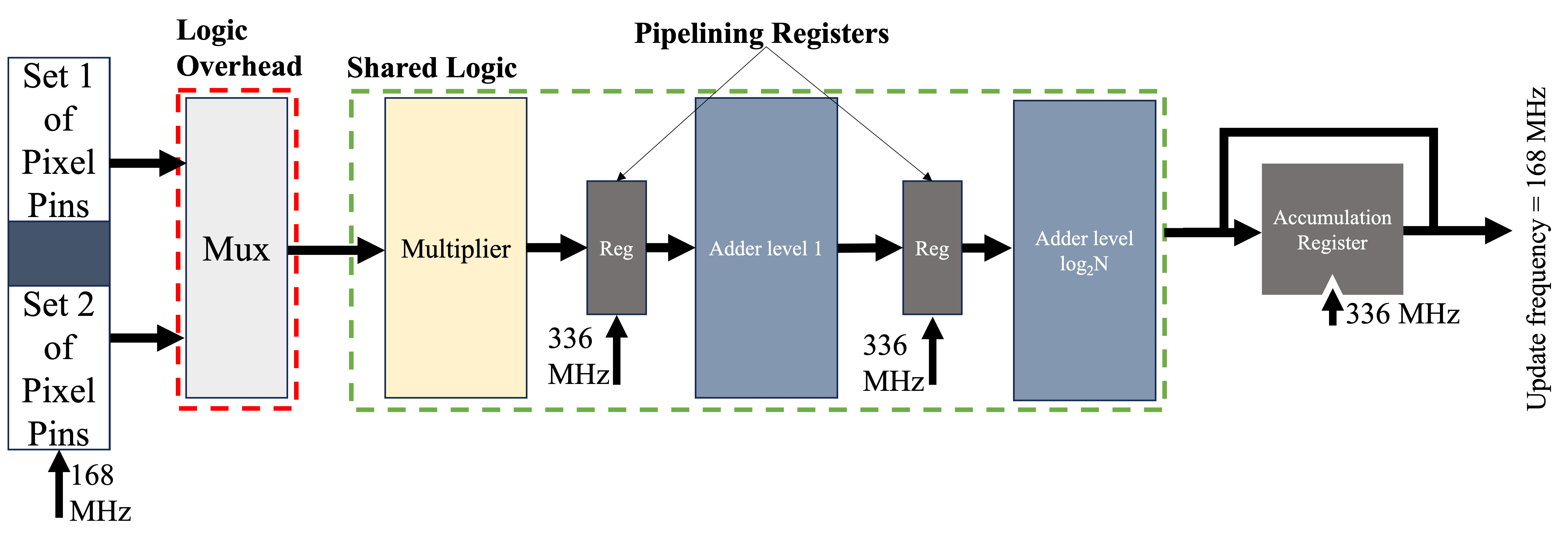}}
\caption{Design optimization.}
\label{opt}
\end{figure}

Illustrated in Figure~\ref{opt} is the logic needed for pipelining and logic sharing in the case of doubling the frequency and sharing the logic twice. Careful consideration must be given to both RTL and logic synthesis to meet the strict timing requirements imposed by the increased frequency needed due to logic sharing.

\subsubsection{Data Width Management}
To save SRAM area, it may be necessary to sacrifice some precision by using smaller weight bit widths, such as FP8 instead of FP16 or FP32. Data width management must be approached carefully to prevent overflow conditions. For example, the data width of the multiplier output should consider both the pixel data width and the weight width. In the accumulator adder tree, the index of each level should determine the data width needed to represent the output at every summation level. Lastly, for the final accumulation register, the number of rows in the frame influences the number of accumulations and should regulate the final width of the accumulation result. By using this approach, the designer prevents overflow and eliminates unnecessary data bits along the matrix multiplication path.

\section{Hardware Design}

\label{sec:RTL_design}
The RTL design is implemented in SystemVerilog HDL. It
provides multiple parameters in the top-level module for
ease of exploring the tradeoff between area and increased
parallel computation. For example, the width of the input
pixels can be adjusted depending on the target pixel matrix size.
Furthermore, the width of the weights stored in the SRAMs can
be adjusted, affecting the precision of the matrix multiply
and the area used by the SRAMs. The dimensions of each operand
can also be adjusted by setting the pixels read per cycle and
the number of weight vectors (i.e., \(K\)). We utilize cocotb~\cite{cocotb}, 
a Python framework for writing simulator-agnostic testbenches, 
and a functional verification tool to simulate our design. 

\section{Determining Necessary Data Widths}
\label{sec:Determining_Necessary_Data_Widths}
Reducing data widths throughout the design is of great interest
as it can greatly reduce the amount of resources needed to implement logic. However, the magnitude of weights and the precision required by our target use cases must be carefully considered. Generally, the width of data determines the scale of representable values and the precision within that scale.

\subsection{Floating Point Format}
Since the majority of our circuit uses floating-point operations, we detail how the floating-point format works and how altering different parameters of the format affects the representable values.
Floating point formats generally have three components: the sign, the exponent (\textit{e}), and the mantissa (\textit{m}). The sign simply determines whether a value is positive or negative. Formats also include an unsigned integer exponent value. A bias (\textit{b}) is applied to this value to allow for the representation of negative exponents. For example, if an exponent is 5 bits wide, a bias of $2^{4} - 1$ may be subtracted from the stored exponent so that values between the orders of $2^{-15}$ and $2^{15}$ can be represented. The last component is referred to as the mantissa and is an unsigned fractional value. This fractional value is added to one and multiplied by $2^{e - b}$ to determine the magnitude of the value represented by the floating point representation. There is also an alternative representation of the format when the exponent is zero. These values are referred to as subnormals. However, we omit this interpretation of the format as we do not implement it in hardware for the sake of reducing complexity.

The sign component is always one-bit wide and can be omitted in the case that all values that need to be represented are either positive or negative. Our use cases require both positive and negative values; thus, this is not a potential point of optimization. The width of the exponent, \textit{e}, effectively controls the scale of representable values. Reducing the width of the exponents means the largest representable value decreases, and the smallest representable non-zero value increases. The precision of the representation is determined by the width of the mantissa, \textit{m}. For example, a 3-bit mantissa can represent the value 1.125. If a floating-point operation results in the value 1.126, this value can be rounded to 1.125. However, if only 2 bits are used, this value must be rounded to 1.25, which is considerably farther from the output value.

\subsection{Floating Point Operator Implementation} 
The multipliers and adders used in the design are based on the IEEE 754 floating-point specification. However, we do not add support for infinity, NaN, or subnormals to reduce the complexity of the operators.
The multipliers can be simplified due to one of the operands being an integer. As a result,
the multiplier logic consists of an integer multiply, taking the mantissa of the weight and the full input value as operands. The leading one of the resulting product is determined through a tree of multiplexers, the output of which is fed into a shifter and rounded using round-to-even policy to produce the mantissa of the final output. The output of the multiplexer tree is also added to the exponent of the weight to determine
the final output's exponent value. Finally, since the input is an unsigned integer, the sign of the weight is passed through as the sign of the final output.

The adders use a comparator and multiplexers to determine which input has the largest magnitude. The mantissas are then normalized by rightward shifting the smaller magnitude number by the difference of the exponents. If the signs of the inputs match, then the normalized mantissas are added. Otherwise, the mantissa of the smaller magnitude number
is subtracted from the other mantissa. In either case, the leading one of the sum is located to determine the new output exponent. The one's position is also used to shift the sum before being rounded using round-to-even policy, resulting in the final output's mantissa. The sign of the larger magnitude input is the sign of the final output.

\subsection{Adjusting Weight Width}

Due to the SRAM having the largest area requirements of all the components in the design, this is a crucial point of optimization. By taking into account how we intend to use matrix multiplication at the detector ASIC, we can determine the minimum number of bits necessary. For example, when using PCA to compress data from the pixel array, we can determine that all weights will be values between -1 and 1, due to the encoding matrix being orthonormal. As a result, no exponent larger than 0
needs to be represented, meaning the bias can be set to  $2^{e} - 1$. With this new bias, the weights need one less bit to
represent the weight range.

\begin{figure}
\centering
\includegraphics[width=.6\columnwidth]{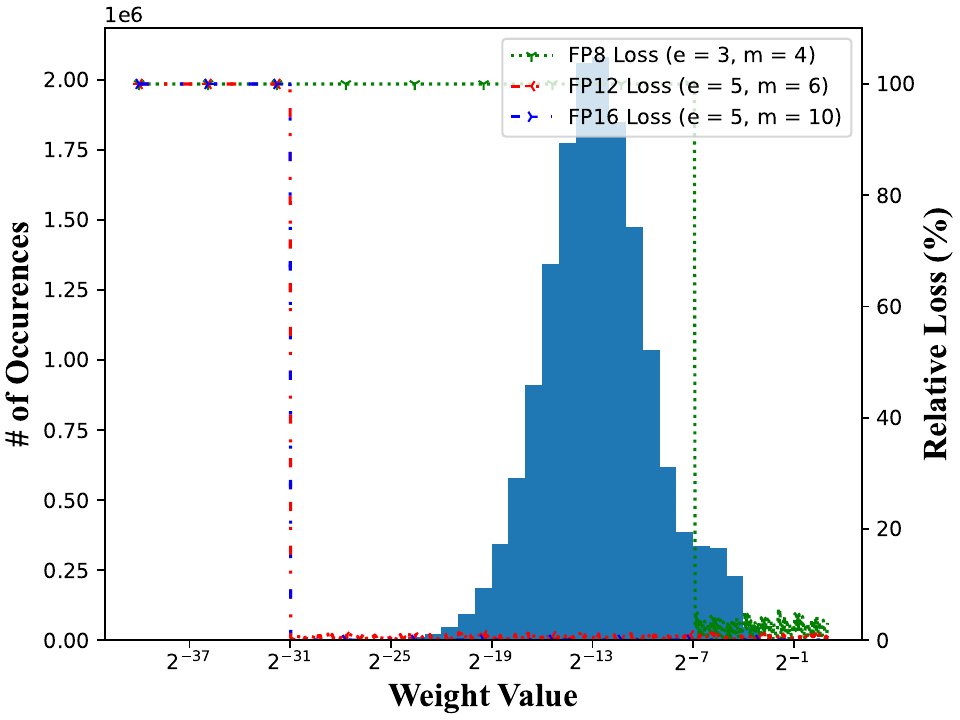}
\caption{The histogram illustrates the frequency of various weight values in an X-ray ptychography dataset. The bars on the left y-axis represent the number of weights across different powers of two. Additionally, there are lines indicating different floating-point formats, which correspond to the relative error (shown on the right y-axis) that occurs when casting a value from FP64 to the specified format. It is important to note that the FP12 weight representation encompasses the required range and only introduces a minor relative error compared to FP16.}
\label{fig:weight_hist}
\end{figure}

However, this change has not resulted in any resource savings, which will only be realized by reducing the width of the different floating-point components. There are some commonly used 8-bit floating point formats where \textit{e} is set to 3 
and \textit{m} set to 4, but we must determine suitability by inspecting the weights used in our target use cases. To do this, we run PCA
on ptychography datasets and inspect the generated weight matrix. Figure~\ref{fig:weight_hist} shows a histogram of the concentration of
weights by value. The bars correspond to the left y-axis, representing the number of weights between different powers of two.
There are also lines for different floating point formats, which correspond to the amount of relative error (right y-axis) introduced by
casting a given value when casting from FP64 to the specified format.
As shown, FP8, which has $e=3$ and $m=4$, experiences extreme losses for most weights, due to them being less than the lowest magnitude representable by FP8.
These values end up being rounded to 0, resulting in 100\% relative error compared to their original format.
This prompts us to develop our own custom format that can adequately cover the entire range of weights.

With \textit{e} set to 5, we can comfortably represent most of the weight matrices (i.e., the lowest exponent would be $-2^{5} - 1 = -31$).
When determining the value \textit{m}, we could set it to 4, just as with the 8-bit format described above, resulting in a 10-bit
format, but we found the precision of this 10-bit format to be a significant deviation from FP16, which has $e=5$ and $m=10$; thus, we developed a 12-bit format with \text{m} set to 6 to improve precision. We found that this format, which we call FP12 ($e=5$ and $m=6$) , still provided significant area savings while producing similar errors to our original 16-bit format.

Note that while the weights now have narrower widths, we do not necessarily change the widths of the output of the multipliers.
Since our input pixel values are unsigned 12-bit integers and our largest weight is of value 1, we want to be able to represent
the same values as the pixel values in our output format. It turns out the products can all be represented in a standard 16-bit
floating point format with only a slight loss in precision. The reasoning behind this is that the range of the floating point
operand is $[-1,1]$ while the range of the integer operand is $[0, 2^{12} - 1]$, meaning the range of the product is $[-2^{12} + 1, -2^{12} + 1]$.
FP16 can represent number of magnitude larger than $2^{12}$, justifying the need to set $e=5$ for the product floating point format. Furthermore,
$m=10$ is chosen since an input value of high magnitude such as $2^{12} - 1$ (i.e. 0b1111111111111111) will require an 11-bit mantissa to represent.
While using 10-bits introduces loss, these extreme values are uncommon in our workloads thus the loss is negligible.

\subsection{Adjusting Width of Accumulation Output}
While adjusting the width of the weights has resulted in significant area savings, adjusting the accumulation width is motivated
more by avoiding overflow. If we consider the max pixel value of $2^{12} - 1$, weights are between values -1 and 1, and that
our design processes a 12\texttimes168 frame, the largest magnitude number that our accumulator can produce is $2^{12} \times 12
\times 168 = 8255520$, which is greater than $2^{22}$. As a result, using IEEE half precision will result in an overflow, since
it can only represent exponents between -15 and 15. If a value goes beyond these bounds, it will result in overflow. To account
for this, we simply widen the output of the last level of adders in the accumulators to 17-bit values with \textit{e} set to 6,
extending the range of exponents to -31 to 31. However, this overflow does not occur when calculating the partial
sum using the adder tree, so the adder tree is entirely composed for 16-bit adders.

\begin{figure}
\begin{subfigure}[b]{.49\columnwidth}
\includegraphics[width=\columnwidth]{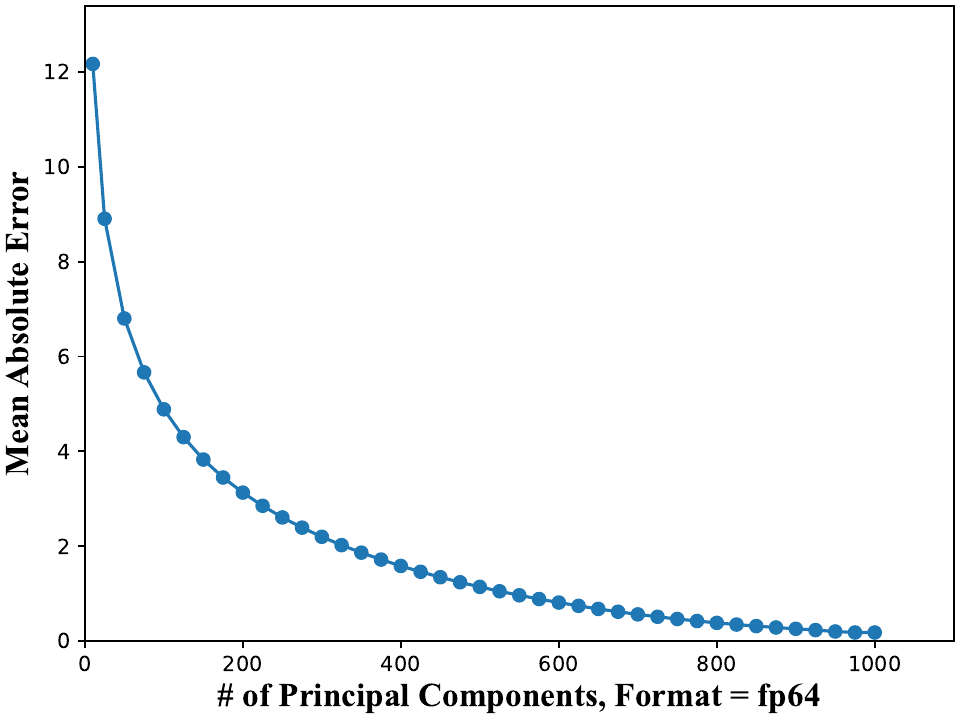}
\caption{}
\label{fig:increasing_k}
\end{subfigure}
\hfill
\begin{subfigure}[b]{.49\columnwidth}
\includegraphics[width=\columnwidth]{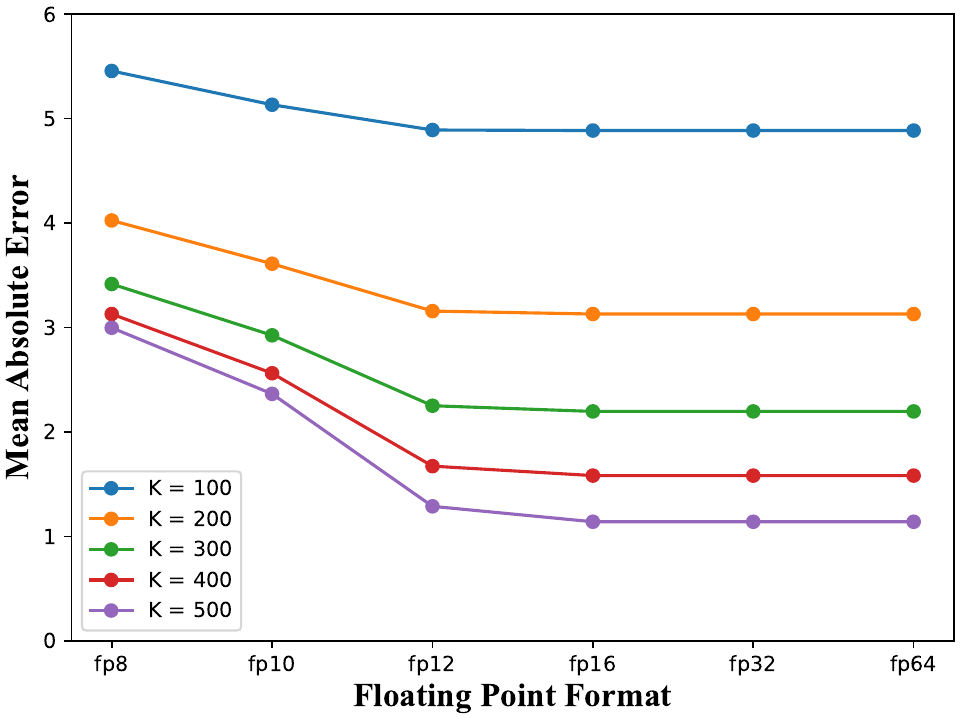}
\caption{}
\label{fig:increasing_width}
\end{subfigure}
\caption{Mean absolute error vs. (a) K, (b) floating-point format.}
\label{errors}
\end{figure}

Figure~\ref{errors} illustrates how the mean absolute error of the PCA algorithm, used as a case study, changes with the number of principal components and the floating-point format employed for weight representation. It is important to note that when using floating-point formats larger than FP12 and with the number of principal components exceeding K = 200, there is only a minimal improvement in the mean absolute error.

\section{Logic Synthesis}
\label{sec:Logic_Synthesis}
Logic synthesis was performed on a \SI{28}{\nm} CMOS technology. Various partitioning schemes and compression ratios were implemented as part of a feasibility study for the current design. The synthesis process considered a pixel array of size 192 \texttimes\ 168, with pixel data width of 12. Two design approaches were implemented. The first approach uses a constant data width of 16 bits for the weights, the output from the multiplier, and the final output of the accumulator. However, this method is not optimal in cases of overflow. The second approach adjusts the data width as needed, starting with 12-bit weights, a 16-bit output from the multiplier, and a 17-bit final accumulation output. This second method effectively saves area and prevents overflow; however, it sacrifices some precision.

\subsection{Constant Data Width Results}
In this approach, the weights were represented using IEEE 754 FP16 format. Rows 1-6 in Table~\ref{tab:table1} summarize the various scenarios considered during the first design approach. It is evident from the table that the SRAM area remains constant when the value of \textit{K} and the weight width are fixed. In contrast, the logic area not only depends on \textit{K} but also varies with different frequencies. The SRAM taken into account in rows 1-6 is the  \SI{28}{\nm} 168\texttimes128 SRAM, which area is 39.345~\textmu m \texttimes\ 166.17~\textmu m, and it was chosen such that each SRAM 128-bit word stores 8 16-bit weights. The compression ratio (CR) in the table is calculated according to
\begin{equation}
\text{CR} = \frac{\text{Raw bits}}{\text{Compressed bits}} 
= \frac{192 \times 168 \times 12}{K \times RW}
\end{equation}
where RW is the final output width.

\subsection{Optimized Data Width Results}
In the second approach, custom FP12 weights are considered, allowing the use of 672\texttimes144 SRAM, which is more efficient in terms of area. The 672-word SRAM compared to the 168-word one benefits from quadrupling the frequency, as we can read from it four times faster, and with quadrupling its depth, it can hold four times more weights, which reduces the total number of SRAMs in the design, and makes the SRAMs more area-efficient. The dimensions of the 28 nm 672\texttimes144 SRAM are 
61.99~\textmu m \texttimes\ 355.75~\textmu m.
Rows 7-9 in Table~\ref{tab:table1} summarize the area and compression ratio results for the second design approach. The compression ratios are expected to be smaller than those of the first approach as the final output width of the accumulator increased by one bit. Table~\ref{tab:table1} shows the power densities of various designs, all of which are below 0.5 W/mm$^2$, significantly lower than the modern design system constraints of 1-2 W/mm$^2$ \cite{rasheedi2025high, abdelzaher2025hybrid}.

Timing closure was achieved in synthesis, with a worst positive slack of at least 10\% of the clock period across all tested frequencies. It is also important to note that the reported synthesis areas are expected to increase by 25–30\% after placement and routing.

\begin{table}[t]
\centering
\caption{The performance of various design scenarios is outlined below. Rows 1-6 in the table represent designs utilizing a constant FP16 width for weights, multipliers, and accumulators. In contrast, rows 7-9 showcase a design that optimizes parameters: using FP12 for weights, FP16 for multipliers, and FP17 for accumulators.\label{tab:table1}}
\smallskip
\resizebox{\textwidth}{!}{%
\begin{tabular}{l
>{\centering\arraybackslash}p{2cm}
|>{\centering\arraybackslash}p{0.5cm}
|>{\centering\arraybackslash}p{2cm}
|>{\centering\arraybackslash}p{2cm}
|>{\centering\arraybackslash}p{2.5cm}
|>{\centering\arraybackslash}p{2cm}
|>{\centering\arraybackslash}p{2cm}
|>{\centering\arraybackslash}p{1.8cm}
|>{\centering\arraybackslash}p{1.8cm}|
>{\centering\arraybackslash}p{2.5cm}|
>{\centering\arraybackslash}p{1.8cm}}
\hline
 & \textbf{Compression Ratio} & \textbf{K} & \textbf{Frequency (MHz)} & \textbf{Frequency Scaling Factor} & \textbf{Weight Width / Accumulator Width} & \textbf{Logic Area (mm$^2$)} & \textbf{SRAM Area (mm$^2$)} & \textbf{Total Area (mm$^2$)} & \textbf{Output Data Rate (Gbps)} & \textbf{Power Consumption (W)} & \textbf{Power Density (W/mm$^2$)} \\
\hline
1  & 252  & 96  & 168 & 1$\times$ & FP16 / FP16    & 16.10 & 15.06 & 31.16   & 1.53 & 4.22 & 0.13 \\
2  & 252  & 96  & 336 & 2$\times$ & FP16 / FP16    & 8.26  & 15.06 & 23.32   & 1.53 & 6.09  & 0.26\\
3  & 252  & 96  & 672 & 4$\times$ & FP16 / FP16     & 3.99  & 15.06 & 19.05   & 1.53 & 6.39  & 0.33\\
4 & 126  & 192 & 168 & 1$\times$ & FP16 / FP16    & 32.44 &30.12 & 62.56   & 3.07 & 8.47  & 0.13\\
5 & 126  & 192 & 336 & 2$\times$ & FP16 / FP16   & 15.03 & 30.12 & 45.15   & 3.07 & 12.12  & 0.26\\
6 & 126  & 192 & 672 & 4$\times$ & FP16 / FP16   & 8.12  & 30.12 & 38.24   & 3.07 & 12.8  & 0.33\\
7 & 237 & 96  & 672 & 4$\times$ & FP12 / FP17  & 3.29 & 8.46  & 11.75 & 1.63 & 4.41 & 0.37\\
8 & 158  & 144 & 672 & 4$\times$ & FP12 / FP17  & 4.85 & 12.7 & 17.55  & 2.44 & 6.58 & 0.37\\
9 & 119  & 192 & 672 & 4$\times$ & FP12 / FP17 & 6.17 & 16.93 & 23.10  & 3.26 & 8.61 & 0.37\\
\hline
\end{tabular}%
}
\end{table}

\section{Physical Implementation}
\label{sec:Physical_Implementation}
\begin{figure}[]
\centerline{\includegraphics[scale = 0.5]{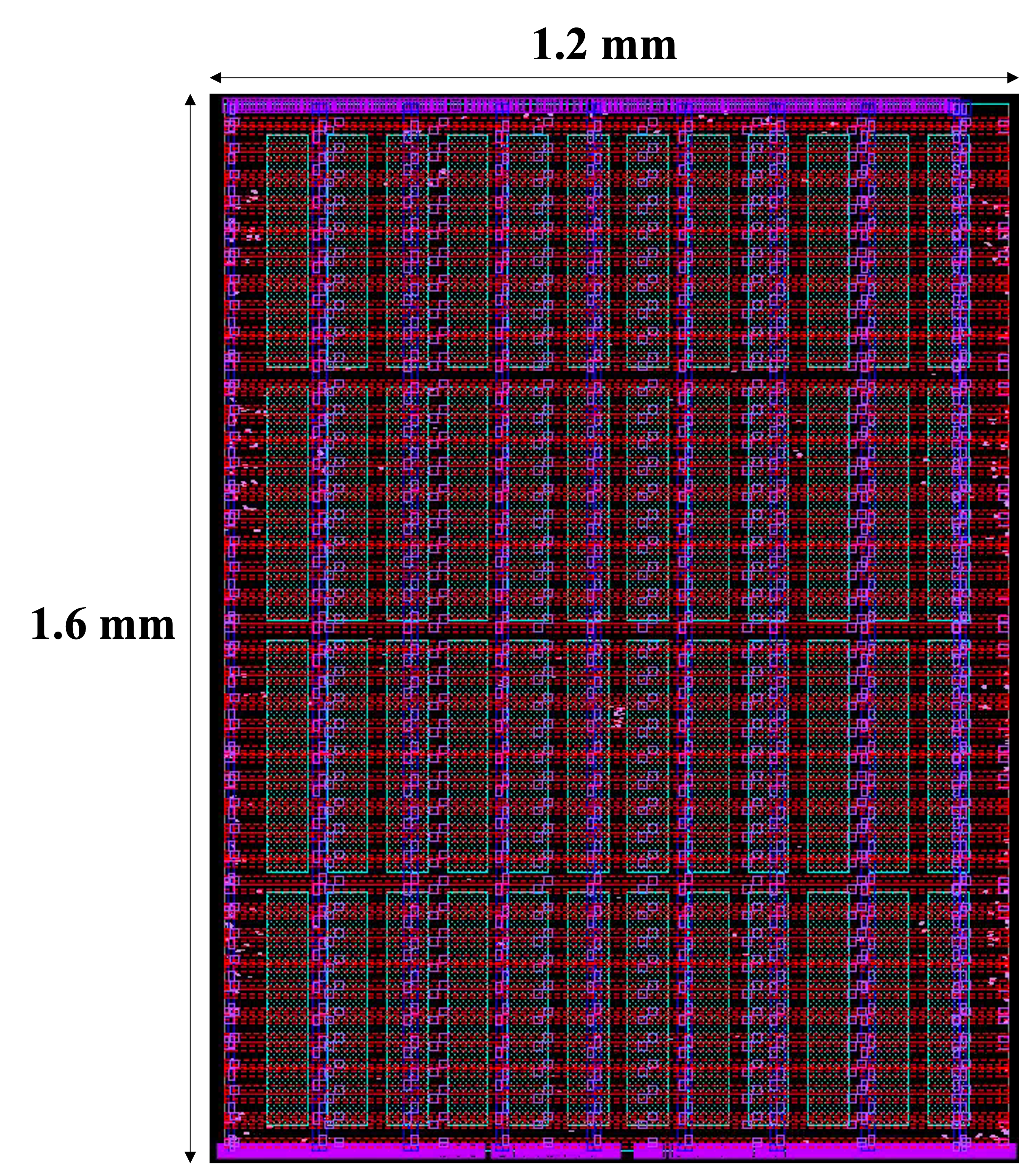}}
\caption{Physical implementation of a 12-column compressor block (i.e., design in row 9 of Table~\ref{tab:table1}). It features pipelining and 4\texttimes\ logic sharing, which area is 1.2 mm \texttimes\ 1.6 mm = 1.92 mm$^2$. This block consists of 48 SRAMs (672 \texttimes\ 144), which totals 4.64 Mbits.}
\label{pnr}
\end{figure}

Figure~\ref{pnr} illustrates the physical implementation of the design in row 9 in table~\ref{tab:table1} for a 12-column section that features pipelining and 4\texttimes\ logic sharing, with an area of 1.2 mm \texttimes\ 1.6 mm = 1.92 mm$^2$. This design is primarily dominated by SRAMs, as depicted in the figure. The final configuration will consist of 16 replicas of this smaller block, followed by an extra adder to reduce the 16 lanes to 1 lane. Each pixel measures 100~\textmu m \texttimes\ 100~\textmu m, which means the width of the design block should be fixed at 12 \texttimes\ 100~\textmu m = 1.2 mm; however, the height can be adjusted based on the area resources and the target cell utilization. The total power consumption of the design in row 9 (i.e., FP12, K = 192, 672 MHz) is 8.613 W at \SI{672}{\MHz} frequency of operation, excluding the power consumption of the extra adder, with 99\% allocated for switching power and the remainder allocated for leakage power. It is important to note that SRAMs account for 68\% of the overall power consumption of the design.

To reduce the number of lanes feeding into the serializing stage, additional levels of addition are necessary after the accumulation stage. For instance, reducing a design from 16 lanes to a single lane for a K value of 192 and an accumulator width of 17 requires four extra levels of summation, consuming approximately 0.61 mm$^2$ of logic area. However, the area and power values reported in Table~\ref{tab:table1} only account for the matrix multiplication and SRAMs, excluding the logic required for result lane reduction and other components such as the serializer and the analog pixel matrix. 

The design is scheduled to be submitted for tape-out, details of which will be presented in a forthcoming publication.

\section{Discussion}
\label{sec:Discussion}
Table \ref{tab:table1} illustrates the trade-offs between compression accuracy, area, data rate, and power consumption. For instance, designs operating at 4\texttimes\ frequency demonstrate significant area savings at the same compression ratios; however, they consume more power than designs operating at 1\texttimes\ frequency (see Table~\ref{tab:table1} rows 1 and 3). Additionally, designs with very small block partitioning and a high value of K achieve realistic and easier physical implementation while preserving good compression accuracy. However, they require more levels of summation after the accumulators to narrow the data down to one lane. Designers should carefully select the appropriate design strategy based on their area, power, and data rate budget. The optimized data width approach generally requires lower SRAM and logic areas, and avoids overflow, which is why it is more favorable than the fixed data width approach. 

If we consider the case for K = 192, this corresponds to 74.3~Mbits of weight data read from the SRAMs each microsecond. This results in an overall memory bandwidth of \SI{74.3}{Tbps}, which is not feasible with current state-of-the-art off-chip memory technologies (e.g., HBM) due to the enormous number of interconnects required (over 100,000 interconnects in the case of K = 192 and FP12 floating-point weights). This substantial bandwidth significantly impacts the overall power consumption of the system.

In-memory computing is an alternative approach for implementing matrix multiplication operations. In this method, the SRAM cell functions as both a memory element for the weights and a multiplier, significantly reducing the area required for multiplication logic~\cite{10040239}. However, addition and accumulation logic will still be necessary to complete the matrix multiplication process. However, as the in-memory computing memories are fully custom, their implementation is potentially more time-consuming and challenging than our approach, where logic and memory are physically segregated. 

Another target use case we intend to explore in future work is neural network inference. Matrix multiplication is the backbone of many neural networks \cite{skansi2018introduction}. Notably, our architecture approximately implements a fully connected neural network layer, which consists of three operations: matrix multiplication, bias addition, and activation. Extending the design to support these additional operations would be feasible within the given area budget. Implementing bias addition would be straightforward, requiring only an extra addition step after the accumulation output. The resulting value would then be passed to an activation function block, whose complexity would depend on the chosen function. However, the rectified linear unit (ReLU), a widely used activation function \cite{daubechies2022nonlinear}, is relatively simple to implement in hardware. ReLU can be described by the following function.

\begin{equation}
f(x) = \begin{cases}
x & \text{if } x > 0, \\
0 & x \le 0
\end{cases}
\end{equation}

This requires only a multiplexer and a comparator circuit to implement in our architecture. Adding these components would enable data-parallel or model-parallel inference near the detector~\cite{jia2019beyond}, which—similar to PCA-based compression—would reduce bandwidth requirements at the detector interface. This first layer could become part of a multilayer neural network, with subsequent layers implemented in later stages of data processing, such as on an FPGA. The advantage of transmitting compressed data significantly benefits these later stages, where fewer neurons (i.e., reduced network complexity) are required, as they are considerably slower than those in the ASIC.

\section{Conclusion}
\label{sec:Conclusion}
 In this work, we present a detailed design study of a scalable multiply-accumulate ASIC architecture in \SI{28}{\nm} CMOS for real-time on-chip data compression in high-throughput X-ray and electron pixel detectors. Targeting a \SI{1}{\MHz} frame rate from a 192\texttimes168 pixel array, our design addresses the critical bottleneck of off-chip bandwidth by leveraging fixed-length, lossy compression based on matrix multiplication with pre-generated encoding weights. We systematically explore architectural trade-offs, including pipelining, logic sharing, and data width optimization, to meet strict area and timing constraints. Logic synthesis and physical implementation results validate the feasibility of integrating compression within a constrained digital footprint, while maintaining high throughput and efficiency. Area-efficient SRAM configurations and dynamic data width tuning further demonstrate significant resource savings. Our results confirm that online, fixed-length compression is not only viable at scale but also essential to enabling next-generation charge-integrating detectors to fully capitalize on advances in X-ray source brightness and detector frame rates. The final implementation will necessarily balance system-level considerations, cost-benefit analysis, and the end-user science requirements. In future work, we will present a comprehensive study of how K, the magnitude of weights, and the precision affect the loss of information on various experimental techniques
 and demonstrate that this multiply-accumulate architecture can also be used for data reduction (e.g., azimuthal integration), making this architecture broadly applicable. 
 
\acknowledgments
This work was supported by the U.S. Department of Energy, Office of Science, Advanced Scientific Computing Research (ASCR) and Basic Energy Sciences (BES), under Contract No. DE-AC02-06CH11357 at Argonne and DE-AC02-76SF00515 at SLAC. This work utilized resources at the Advanced Photon Source, a U.S. Department of Energy Office of Science User Facility. Work performed at the Center for Nanoscale Materials and Advanced Photon Source, both U.S. Department of Energy Office of Science User Facilities (SUF), was supported by the U.S. DOE, Office of Basic Energy Sciences, under Contract No. DE-AC02-06CH11357. Work at University of Chicago is supported by the Divisions of Chemistry (CHE) and Materials Research (DMR), National Science Foundation, under grant numbers NSF/CHE-1834750 and NSF/CHE-2335833. This work was primarily supported by the AUREIS project (part of Microelectronics Energy Efficiency Research Center for Advanced Technologies (MEERCAT)) and the Morpheus project, supported by DOE BES/SUF's Accelerator and Detector R\&D program.

\bibliographystyle{JHEP}
\bibliography{biblio}

\end{document}